\documentclass[aps,prl,twocolumn,showpacs,superscriptaddress,groupedaddress]{revtex4}  
\usepackage{amssymb}   
\usepackage{amsmath}

\newcommand{\cM}{\begin{cal}M\end{cal}}
\newcommand{\ord}{\begin{cal}O\end{cal}}

\newcommand{\cS}{\begin{cal}S\end{cal}}

\newcommand{\eps}{\epsilon}

\def\beq{\begin{equation}}
\def\eeq{\end{equation}}
\def\bsp#1\esp{\begin{split}#1\end{split}}

\begin{document}

\hspace{7cm} \mbox{IPPP/13/70, DCPT/13/140, ZU-TH 18/13, NSF-KITP-13-176}

\title{The two-loop soft current in dimensional regularization}

\author{Claude Duhr}
\affiliation{Institute for Theoretical Physics, ETH Z\"urich,
  8093 Z\"urich, Switzerland}
\affiliation{Institute for Particle Physics Phenomenology,
University of Durham, Durham, DH1 3LE, U.K.}

\author{Thomas Gehrmann}
\affiliation{Institut f\"ur Theoretische Physik, Universit\"at Z\"urich, 
Winterthurerstrasse 190, 8057 Z\"urich, Switzerland}
\affiliation{Kavli Institute for Theoretical Physics, University of California, 
Santa Barbara, CA 93106, USA}
\date{\today}

\begin{abstract}
The soft current describes the factorization behavior of 
quantum chromodynamics (QCD) scattering amplitudes in the 
limit of vanishing energy of one of the external partons. It is  
process-independent  and can be expanded in a perturbative series in 
the coupling constant. To all orders in the  dimensional regularization
parameter, we compute the two-loop 
correction to the soft current for processes involving two hard partons.  
\end{abstract}

\maketitle

In the limit where one or more massless partons are unresolved, amplitudes 
in quantum field theory 
factorize into lower-point amplitudes with the unresolved partons removed, 
times a universal function that is independent of the details of the hard interaction 
and describes the emission of the unresolved particles.
This factorization property has multiple implications both for the formal study of scattering 
in quantum field theory and for the phenomenology of scattering processes at 
high-energy particle colliders. On the formal side, several conjectures on the high-order 
behavior of perturbation theory and on the all-order structure of the scattering matrix 
are formulated based on insights gained from the limiting behavior in unresolved limits.
In precision applications of perturbation theory to collider phenomenology, 
systematic expansions 
around unresolved limits allow to approximate or reconstruct higher-order coefficients 
in an elegant and computationally efficient manner. 


In particular, the emission of a soft gluon is entirely described by the so-called \emph{soft current}, an operator in color space that encapsulates all the information on the soft emission. 
In this letter we report on the computation of the two-loop soft current in QCD for the emission of a single soft gluon from an amplitude involving two hard colored particles. 

Let us consider the amplitude $|\cM(q,p_1,\ldots,p_n)\rangle$ (as  vector in color space) for a gluon with momentum $q$ in association with $n$ colored particles with momenta $p_i$, $i=1,\ldots,n$, transforming in some irreducible representation of $SU(N)$. The amplitude may depend 
furthermore on an arbitrary number of colorless particles. In the soft limit where the energy of the gluon vanishes, the amplitude factorizes according to
\beq\label{eq:def_soft_current}
\hskip -0.009cm
\langle a|\cM(q,p_1,\ldots,p_n)\rangle \simeq \varepsilon^\mu(q)\,J_\mu^a(q)\,|\cM(p_1,\ldots,p_n)\rangle\,,
\eeq
where $a$ denotes the adjoint color index of the soft gluon and $\varepsilon^\mu(q)$ its polarization vector, and the `$\simeq$' sign indicates that the equality only holds up to the leading term in the expansion in the soft gluon momentum. Equation~\eqref{eq:def_soft_current} defines the (unrenormalized) soft current $J_\mu^a(q)$, which describes the emission of a soft gluon. Both the soft current and the amplitude admit a perturbative expansion,
\beq\bsp
|\cM(p_1,\ldots,p_n)\rangle &\,= \sum_{\ell=0}^\infty|\cM^{(\ell)}(p_1,\ldots,p_n)\rangle\,,\\
J_\mu^a(q) &\, = g_S\,\mu^\eps\,\sum_{\ell=0}^\infty(g_S\,\mu^\eps)^{2\ell}\,J_\mu^{a(\ell)}(q) \,.
\esp\eeq
We work in $D=4-2\eps$ dimensions and $g_S$ denotes the bare QCD coupling constant and $\mu$ is the scale introduced by dimensional regularization. The tree-level soft current is given by the well-known eikonal factor,
\beq
J_\mu^{a(0)}(q) = \sum_{i=1}^nT^a_i\,\frac{p_{i\mu}}{p_i\cdot q}\,,
\eeq
where $T^a_i$ denote the generators of $SU(N)$ of the representation of parton $i$. The one-loop correction was computed in ref.~\cite{Catani:2000pi},
\beq\bsp\label{eq:J1}
J&_\mu^{a(1)}(q) = -\frac{S_\eps}{16\pi^2}\,\frac{1}{\eps^2}\,\Gamma(1-\eps)\Gamma(1+\eps)\,if^{abc}\\
&\,\times \sum_{i\neq j}T_i^b\,T_j^c\,\left(\frac{p_{i\mu}}{p_{i}\cdot q}-\frac{p_{j\mu}}{p_{j}\cdot q}\right)\,\left[\frac{(-s_{ij})}{(-s_{iq})(-s_{qj})}\right]^\eps\,,
\esp\eeq
with $s_{ij} = 2p_i\cdot p_j + i0$ and 
\beq\bsp
S_\eps=(4\pi)^\eps\,\frac{\Gamma(1+\eps)\Gamma(1-\eps)^2}{\Gamma(1-2\eps)}\,.
\esp\eeq
We emphasize that the soft current is not a scalar quantity, but an operator in color space, i.e., it acts non-trivially on the color indices of the hard amplitude. While at tree and one-loop level, all color operators involve at most two hard partons, starting from two loops new color structures may appear that connect up to three hard partons~\cite{dipole}. Non-trivial contributions from these new 
color structures to the soft current cannot be excluded. 

If we focus on processes with only two hard colored particles, the color structure of the soft current drastically simplifies. Indeed, color conservation in the hard amplitude implies that the two hard partons must transform in complex conjugate representations. The color structure is then most conveniently described by color-ordered helicity amplitudes,
\beq
|\cM(q,p_1,p_2)\rangle = T^a_{i_1i_2}\,A(q,p_1,p_2)\,,
\eeq
where the color-ordered amplitude $A(q,p_1,p_2)$ depends on the  helicities and 
 momenta of the colored particles, but not on their color. In the limit where the gluon becomes soft, the color-ordered amplitude factorizes,
\beq
A(q,p_1,p_2)\simeq g_S\,\mu^{\eps}\,\cS_\pm(q)\,r_{soft}(q)\,A(p_1,p_2)\,.
\eeq
The helicity dependence of the soft emission is entirely encoded into the tree-level soft function,
\beq
\cS_+(q) = \sqrt{2}\,\frac{\langle12\rangle}{\langle1q\rangle\langle q2\rangle}{\rm~~and~}
\cS_-(q) = -\sqrt{2}\,\frac{[12]}{[1q][q2]}\,.
\eeq
Here $\langle ij\rangle$ and $[ij]$ denote the usual spinor products, related to the Mandelstam invariants by $s_{ij} = \langle ij\rangle\,[ji]$. Quantum corrections to the soft emission are helicity-independent 
and expressed in the scalar function 
\beq\bsp
r&_{soft}(q) = \\
&1+\sum_{\ell=1}^\infty\left\{g_S^2\,\mu^{2\eps}\,\frac{S_\eps}{16\pi^2}\left[\frac{(-s_{12})}{(-s_{1q})(-s_{q2})}\right]^\eps\right\}^\ell\,r_{soft}^{(\ell)}\,.
\esp\eeq
The coefficients $r_{soft}^{(\ell)}$ are related to the soft current by
\beq\bsp\label{eq:J0Jl}
J^{a(0)}_\mu(q)\,J^{\mu(\ell)}_a(q) &\,= -4\,C_i\,\left(\frac{S_\eps}{16\pi^2}\right)^\ell\\
&\,\times\left[\frac{(-s_{12})}{(-s_{1q})(-s_{q2})}\right]^{1+\ell\eps}\,r_{soft}^{(\ell)}\,,
\esp\eeq
where $C_i$ is the Casimir operator in the representation of the two hard partons. Comparing eq.~\eqref{eq:J0Jl} to eq.~\eqref{eq:J1} and performing the color algebra, we immediately see that the one-loop coefficient is given by
\beq\label{eq:r1soft}
r_{soft}^{(1)} = -N\,\frac{\Gamma(1-\eps)\Gamma(1+\eps)}{\eps^2}\,.
\eeq
The two-loop coefficient was computed in 
ref.~\cite{Badger:2004uk} up to $\mathcal{O}(\eps^0)$ by considering the soft limit of the 
two-loop amplitudes for $\gamma^*\to Q\,\bar Q\, g$~\cite{Garland:2002ak} 
and  $H\to 3$~partons~\cite{Gehrmann:2011aa}. For applications in precision calculations, 
the soft current is to be integrated over the soft phase space (giving 
rise to a double pole in the regularization parameter), and is therefore 
required to $\mathcal{O}(\eps^2)$. 


The two-loop coefficient $r_{soft}^{(2)}$ can be extracted from a given two-loop amplitude involving two hard partons and a gluon. We focus on the $D$-dimensional two-loop amplitude for $\gamma^*\to Q\,\bar{Q}\,g$, interfered with the tree-level amplitude and summed over colors and spins~\cite{Garland:2001tf}. The matrix element is a function of the lightlike momenta $p_1$, $p_2$ and $q$ of the quark pair and the gluon. In the limit where the gluon becomes soft, it factorizes according to,
\beq\bsp\label{eq:QQg_factorization}
\langle&\cM^{(0)}_3|\cM^{(2)}_{3}\rangle \simeq\\
& -g_S^2\,\mu^{2\eps}\,\sum_{\ell=0}^2(g_S^2\,\mu^{2\eps})^\ell\,
\langle\cM^{(0)}_2|J_\mu^{a(0)}J_a^{\mu(\ell)}|\cM^{(2-\ell)}_{2}\rangle\,.
\esp\eeq
The two-loop coefficient $r_{soft}^{(2)}$ can then directly be extracted by expanding in the soft-gluon momentum.

If we denote the virtuality of the photon by $Q^2=(p_1+p_2+q)^2$, then (up to some overall power of $Q^2$) the matrix element $\langle\cM^{(0)}_3|\cM^{(2)}_{3}\rangle$ can only depend on the Lorentz-invariant dimensionless ratios
\beq\label{eq:xyz}
x=\frac{s_{12}}{Q^2}\,,\qquad y=\frac{s_{1q}}{Q^2}\,,\qquad z=\frac{s_{2q}}{Q^2}\,,
\eeq
subject to the constraints 
\beq
x+y+z=1 {\rm~~and~~}0<x,y,z<1\,.
\eeq 
Without loss of generality, we set $Q^2=1$ in the following.
The soft limit is then approached when both $y$ and $z$ tend to zero at the same rate. Our goal is thus to expand the matrix element into a power series in $y$ and $z$ while keeping the dependence of the coefficients on the dimensional regulator $\eps$ exact. The leading term of the expansion then corresponds to the right-hand side of eq.~\eqref{eq:QQg_factorization}. 

The two-loop amplitude for $\gamma^*\to Q\,\bar{Q}\,g$ can be written as a linear combination of scalar four-point master integrals with one external massive leg~\cite{Gehrmann:1999as,Gehrmann:2000zt}. In the following we denote the master integrals collectively by $F_i(y,z;\eps)$. 
The master integrals themselves satisfy a system of coupled differential equations that can schematically be written as
\beq\bsp\label{eq:diffeqs}
\frac{\partial}{\partial y}F_i(y,z;\eps) &\,= A^y_{ij}(y,z;\eps)\,F_j(y,z;\eps)\,,\\
\frac{\partial}{\partial z}F_i(y,z;\eps) &\,= A^z_{ij}(y,z;\eps)\,F_j(y,z;\eps)\,,
\esp\eeq
where $A^k_{ij}(y,z;\eps)$, $k\in\{y,z\}$, are rational functions of $y$, $z$ and $\eps$. 
Solutions to eqs.~\eqref{eq:diffeqs} valid to all orders in $\eps$ are only available in a few special cases~\cite{Gehrmann:1999as}. Laurent expansions in $\eps$ 
were obtained for all master integrals 
up to $\ord(\eps^0)$  in terms of harmonic polylogarithms and their two-dimensional generalization~\cite{Gehrmann:2000zt}.
These results yield the two-loop amplitude for $\gamma^*\to Q\,\bar{Q}\,g$ up to $\ord(\eps^0)$. Expanding the two-dimensional harmonic polylogarithms as power series in $y$ and $z$ immediately reproduces the known result for $r_{soft}^{(2)}$ up to $\ord(\eps^0)$~\cite{Badger:2004uk}.

To obtain the two-loop coefficient $r_{soft}^{(2)}$ to all orders in $\eps$, we 
return to the differential equations~\eqref{eq:diffeqs} and construct for each master integral
a power series solution in $y$ and $z$ close to the origin $(y,z)=(0,0)$ in the $(y,z)$ plane. The differential equations may, however, have poles whenever $y$ or $z$ vanish, translating into branching points for the master integrals starting from points where one of the two expansion parameters is zero. In other words, the solutions to eq.~\eqref{eq:diffeqs} are not meromorphic in a neighborhood of the origin of the $(y,z)$ plane, and so we cannot make a simple Laurent series ansatz in $y$ and $z$ for the master integrals.
The correct ansatz for each master integral rather takes the form
\beq\label{eq:ansatz1}
F_i(y,z;\eps) = 
\sum_{m,n=0}^2\,y^{-m\eps}\,z^{-n\eps}\,f_{i,mn}(y,z;\eps)\,,
\eeq
where the $f_{i,mn}(y,z;\eps)$ are meromorphic in a neighborhood of the origin. As such they admit a Laurent series expansion,
\beq\label{eq:ansatz2}
f_{i,mn}(y,z;\eps) = \sum_{k=r_y}^\infty\sum_{l=r_z}^\infty\,c_{i,mn}^{kl}(\eps)\,y^k\,z^l\,,
\eeq
where the $c_{i,mn}^{kl}(\eps)$ are meromorphic functions of $\eps$ and $r_y,r_z\in\mathbb{Z}$. Inserting the ans\"atze~\eqref{eq:ansatz1} and~\eqref{eq:ansatz2} into the differential equations~\eqref{eq:diffeqs} and expanding the functions $A^k_{ij}(y,z;\eps)$ into a Laurent series around the origin, we obtain a linear system for the coefficients $c_{i,mn}^{kl}(\eps)$. The solution to the linear system then provides us with the desired (truncated) Laurent series solution close to the origin. 
Since we did not expand in $\eps$ at any stage, the solutions for the coefficients are exact in $\eps$.

Since eq.~\eqref{eq:diffeqs} is a system of first-order differential equations, 
 one coefficient per master integral is related to boundary conditions, and thus 
not fixed by solving the linear system. In many cases the boundary 
condition can either be inferred because the homogeneous solution 
does not take the form~\eqref{eq:ansatz1} or by requiring 
consistency when solving the linear system.
In the remaining cases, an explicit integral representation for the leading term in the soft expansion of the integral can be derived using the technique of \emph{expansion by regions}~\cite{Pak:2010pt}, which allows one to compute asymptotic expansions of Feynman integrals when some of the external parameters are small. The initial condition for a given master integral can then be fixed by requiring the leading term of the general solution to the differential equation to agree with the result obtained from expansion by regions.

We have applied this strategy to obtain the first few terms in the expansion of all the master integrals in a neighborhood of the origin. 
We have checked that in all cases our results agree, after expanding the coefficients $c_{i,mn}^{kl}(\eps)$ in $\eps$, with the soft expansion of the known results for the master integrals~\cite{Gehrmann:2000zt} in terms of two-dimensional harmonic polylogarithms. 

Having obtained the expansions of all master integrals in the soft limit, we can immediately extract the function $r_{soft}^{(2)}$ from the two-loop matrix element for $\gamma^*\to Q\,\bar{Q}\,g$. 
After inserting the expansions of the master integrals in the soft limit, we see that the leading term of the expansion takes the form
\beq\label{eq:soft_two_loop_structure}
\langle\cM^{(0)}_3|\cM^{(2)}_3\rangle \simeq \sum_{k=0}^2{A_k(\eps)}\,{y^{-1-k\eps}\,z^{-1-k\eps}} \,,
\eeq
in agreement with eq.~\eqref{eq:QQg_factorization}. Comparing eq.~\eqref{eq:soft_two_loop_structure} to eq.~\eqref{eq:QQg_factorization}, we can read off the result for the two-loop coefficient $r_{soft}^{(2)}$. Amazingly, we observe that only planar master integrals contribute to $r_{soft}^{(2)}$. We obtain,
\beq\label{eq:r2soft}
r_{soft}^{(2)} = N\,N_f\,R_1(\eps)+N^2\, R_2(\eps)\,,
\eeq
with
\begin{widetext}
{\allowdisplaybreaks
\begin{eqnarray}\label{eq:R1}
R_1(\eps) &=&\frac{ 2\,\Gamma(-2\epsilon)}{(1+\epsilon)\,\Gamma(4-2\epsilon )}\,\, \frac{ \Gamma (1-2 \epsilon )^2\,\Gamma (1+2 \epsilon )^2}{\Gamma (1-\epsilon )^2 \Gamma (1+\epsilon)^2}
\,\left[3\,\frac{\Gamma (1-\epsilon )\Gamma (1-2 \epsilon )}{\Gamma (1-3 \epsilon )}-\frac{\left(1+\eps^3\right) }{\epsilon ^2\,(1+\eps)}\,\frac{\Gamma (1-2 \epsilon )^2}{ \Gamma (1-4 \epsilon )}\right] \,,\\
\label{eq:R2}
R_2(\eps)&=& \frac{\Gamma (1-2 \epsilon )^3 \Gamma (1+2 \epsilon)^2 }{6\,\epsilon ^4\, \Gamma (1-\epsilon ) \Gamma (1+\epsilon )^2\Gamma(1-3\eps)}\,\Bigg\{(1+4 \epsilon )\, _4F_3(1,1,1-\epsilon ,-4 \epsilon ;2,1-3 \epsilon ,1-2 \epsilon ;1)\\
\nonumber&-&6\epsilon\,\big[\psi(1-3 \epsilon )+\psi(1-2 \epsilon )-\psi(1-\epsilon )-\psi(1+\epsilon)\big]+\frac{\left(14 \epsilon ^3+4 \epsilon ^2+5 \epsilon -3\right)}{2 (1+\epsilon) (3-2 \epsilon) (1-2 \epsilon )}\Bigg\}\nonumber \\
&+&\frac{(1+4 \epsilon )}{3\,\eps^4 \,(1+2 \epsilon ) }\frac{\Gamma (1-2 \epsilon )^4 \Gamma (1+2 \epsilon)^2 }{\Gamma (1-\epsilon )^2 \Gamma (1+\epsilon)^2\Gamma(1-4\eps)}\,\Bigg\{2\, _3F_2(1,-2 \epsilon ,2 \epsilon +1;1-\epsilon ,2 \epsilon +2;1)\nonumber 
\\
\nonumber&-&\frac{ \Gamma (1+\epsilon ) \Gamma (1-2 \epsilon ) }{\Gamma (1-\epsilon )}\, _3F_2(-2 \epsilon ,\epsilon +1,2 \epsilon +1;1-\epsilon ,2 \epsilon +2;1)+\frac{(1+2\eps)\left(6 \epsilon ^4+13 \epsilon ^3-16 \epsilon ^2-38 \epsilon +3\right)}{4(1+4\eps) (1+\epsilon ) (3-2 \epsilon) (1-2 \epsilon )}\Bigg\}\,,
\end{eqnarray}}
\end{widetext}
where $\psi(z)$ denotes the digamma function
and ${_pF_q}$ are the generalized hypergeometric functions.
Equations~(\ref{eq:r2soft} -- \ref{eq:R2}) are the main results of this paper. We stress that our results are valid to all orders in $\eps$. The hypergeometric functions can be expanded into a Laurent series in $\eps$ using standard techniques~\cite{hypexp}, and it is easy to see that 
the Laurent expansion of $r_{soft}^{(2)}$ only involves multiple zeta values to all orders in $\eps$. In particular, the first few orders read explicitly
\beq\bsp
&R_1(\eps) = \frac{1}{6 \epsilon ^3}+\frac{5}{18 \epsilon ^2}+\frac{1}{\epsilon }\left(\frac{1}{3}\,\zeta _2+\frac{19}{54}\right)-\frac{8 }{3}\,\zeta _3\\
&\,+
\frac{5 }{9}\,\zeta _2+\frac{65}{162} +
\left(-6\, \zeta _4-\frac{40 }{9}\,\zeta _ 3-\frac{8}{27}\,\zeta _ 2+\frac{211}{486}\right) \epsilon\\
&\,
+\bigg(-32\, \zeta _5-\frac{16}{3}\, \zeta _ 3\, \zeta _ 2-10\, \zeta _ 4-\frac{287}{27}\, \zeta _ 3-\frac{151 }{81}\,\zeta _2\\
&\,+\frac{665}{1458}\bigg) \epsilon ^2 
+\ord(\eps^3) \,,\\
&R_2(\eps) = \frac{1}{2 \epsilon ^4}-\frac{11}{12 \epsilon ^3}+\frac{1}{\eps^2}\left(\frac{3 }{2}\,\zeta _2-\frac{67}{36}\right)+\frac{1}{\eps}\left(\frac{1}{2}\,\zeta _3\right.\\
&\,\left.-\frac{11}{6}\,\zeta _2-\frac{193}{54}\right)+\frac{29}{4}\,\zeta _4+\frac{44}{3}\, \zeta _3
-\frac{67 }{18}\,\zeta _2 -\frac{571}{81}\\
&\,
+\left(-\frac{37}{2}\,\zeta _ 5+4\, \zeta _ 3\, \zeta _ 2+33\, \zeta _ 4+\frac{268 }{9}\,\zeta_3-\frac{166 }{27}\,\zeta _ 2\right.\\
&\,\left.-\frac{3410}{243}\right) \epsilon
+\bigg(-\frac{451 }{12}\,\zeta _ 6-\frac{29 }{2}\,\zeta _ 3^2+176\, \zeta _ 5+\frac{88 }{3}\,\zeta _ 2\, \zeta _3 \\ &\, +67 \zeta _4+\frac{1679 }{27}\,\zeta _ 3-\frac{1007 }{81}\,\zeta _ 2-\frac{20428}{729}\bigg) \epsilon ^2+\ord(\eps^3)\,.
\esp\eeq
We checked that our result agrees with the result of ref.~\cite{Badger:2004uk} up to $\ord(\eps^0)$. 



While our result completely describes the emission of a soft gluon from a two-loop QCD amplitude with two hard partons, it does not seem trivial to extend the result to higher-point amplitudes. Indeed, starting from two loops, color-correlations among three hard partons can no longer be excluded. Such correlation are however not covered by eq.~\eqref{eq:J0Jl}, which is only valid for two hard partons. This interesting problem, which is connected to the possibility of color-correlations among four hard massless partons in the three-loop soft anomalous dimension~\cite{dipole}, needs further investigation.

An immediate application of our result are the third-order QCD corrections (N$^3$LO) 
to the inclusive production of a Higgs boson or a gauge boson in hadron-hadron collisions. Especially 
in the case of Higgs boson production, these corrections are very much demanded to lower the 
theory uncertainty on the dominant gluon fusion contribution to Higgs production to a level allowing 
precision studies of the Higgs boson. Several ingredients to these corrections have been derived 
previously~\cite{higgsn3lo}, and the two-loop soft current derived here was a crucial missing 
ingredient. 

To summarize,  we have computed the two-loop corrections to the QCD soft-current for the emission of a soft gluon from an amplitude involving at most two hard partons in addition to the soft gluon. Our result is valid to all orders in dimensional regularization and is characterized by a remarkably compact functional form involving only simple hypergeometric functions that can easily be expanded into a Laurent series in $\eps$ with multiple zeta values as coefficients.

{\bf Acknowledgements:} We thank Lance Dixon and Matthias Neubert for clarifying discussions. 
TG likes to thank the Kavli Institute for Theoretical Physics (KITP) at UC Santa 
Barbara for hospitality while this work was completed.
This research is supported in part by the Swiss National Science Foundation (SNF) under 
contract 200020-138206, by the European Commission through the ERC grant ``IterQCD'' and
the``LHCPhenoNet" Initial Training Network PITN-GA-2010-264564
 and by the National Science Foundation under Grant NSF PHY11-25915.
After this paper was finalized, we became aware of independent work~\cite{lz} 
by Ye Li and Hua Xing Zhu, deriving the two-loop soft gluon current to ${\cal O}(\epsilon^2)$, in 
full agreement with our results.



\begin{thebibliography}{99}

\bibitem{Catani:2000pi} 
  S.~Catani, M.~Grazzini,
  Nucl.\ Phys.\ B {\bf 591}, 435 (2000).

\bibitem{dipole}
E.~Gardi, L.~Magnea,
JHEP {\bf 0903}, 079  (2009);
NuovoCim. C {\bf 32N5-6}, 137 (2009);
%
T.~Becher, M.~Neubert,
Phys.\ Rev.\ Lett.\ {\bf 102}, 162001 (2009) 162001;
JHEP {\bf 0906}, 081  (2009).

  
\bibitem{Badger:2004uk}
  S.D.~Badger, E.W.N.~Glover,
  JHEP {\bf 0407}, 040 (2004).
  


\bibitem{Garland:2002ak} 
  L.W.~Garland, T.~Gehrmann, E.W.N.~Glover, A.~Koukoutsakis, E.~Remiddi,
  Nucl.\ Phys.\ B {\bf 642}, 227 (2002).
  
\bibitem{Gehrmann:2011aa} 
  T.~Gehrmann, M.~Jaquier, E.W.N.~Glover, A.~Koukoutsakis,
  JHEP {\bf 1202}, 056 (2012).

  
\bibitem{Garland:2001tf}
  L.W.~Garland, T.~Gehrmann, E.W.N.~Glover, A.~Koukoutsakis, E.~Remiddi,
  Nucl.\ Phys.\ B {\bf 627}, 107 (2002).
  
\bibitem{Gehrmann:1999as}
  T.~Gehrmann, E.~Remiddi,
  Nucl.\ Phys.\ B {\bf 580}, 485  (2000).
  
\bibitem{Gehrmann:2000zt}
  T.~Gehrmann, E.~Remiddi,
  Nucl.\ Phys.\ B {\bf 601}, 248 (2001);
  %
  Nucl.\ Phys.\ B {\bf 601}, 287 (2001).
  
\bibitem{Pak:2010pt}
M.~Beneke, V.A.~Smirnov, Nucl.\ Phys.\ {\bf B522}, 321 (1998);
%
V.A.~Smirnov, Phys.\ Lett.\ {\bf B465}, 226 (1999);
Springer Tracts Mod.\ Phys.\ {\bf 177}, 1 (2002);
%
  A.~Pak, A.~Smirnov,
  Eur.\ Phys.\ J.\ C {\bf 71}, 1626 (2011).

\bibitem{hypexp}
 T.~Huber, D.~Ma\^{\i}tre,
  Comput.\ Phys.\ Commun.\  {\bf 175}, 122 (2006);
  Comput.\ Phys.\ Commun.\  {\bf 178}, 755 (2008).
\bibitem{higgsn3lo}
 P.A.~Baikov, K.G.~Chetyrkin, A.V.~Smirnov, V.A.~Smirnov, M.~Steinhauser,
  Phys.\ Rev.\ Lett.\  {\bf 102}, 212002 (2009);
T.~Gehrmann, E.~W.~N.~Glover, T.~Huber, N.~Ikizlerli, C.~Studerus,
  JHEP {\bf 1006}, 094 (2010);
   C.~Anastasiou, S.~B\"uhler, C.~Duhr and F.~Herzog,
  JHEP {\bf 1211}, 062 (2012);
   M.~H\"oschele, J.~Hoff, A.~Pak, M.~Steinhauser, T.~Ueda,
  Phys.\ Lett.\ B {\bf 721}, 244 (2013);
  C.~Anastasiou, C.~Duhr, F.~Dulat, B.~Mistlberger,
  JHEP {\bf 1307}, 003 (2013);
   S.~B\"uhler, A.~Lazopoulos,
  arXiv:1306.2223.
 \bibitem{lz}
 Ye Li, Hua Xing Zhu,  SLAC-PUB-15732. 
  
\end{thebibliography}
\end{document}